# Electroforming Free Controlled Bipolar Resistive Switching in Al/CoFe$_2$O$_4$/FTO device with Self-Compliance Effect

Sandeep Munjal and Neeraj Khare[*]

Department of Physics, Indian Institute of Technology Delhi, Hauz Khas, New Delhi-110016, India.

**Abstract**

Controlled bipolar resistive switching (BRS) has been observed in nanostructured CoFe$_2$O$_4$ films using Al(aluminum)/CoFe$_2$O$_4$/FTO(fluorine-doped tin oxide) device. The fabricated device shows electroforming-free uniform BRS with two clearly distinguished and stable resistance states without any application of compliance current (CC), with a resistance ratio of high resistance state (HRS) and low resistance state (LRS) $> 10^2$. Small switching voltage ($< 1$ volt) and lower current in both the resistance states confirms the fabrication of low power consumption device. In the LRS, the conduction mechanism was found to be of Ohmic in nature, while the high-resistance state (HRS/OFF state) was governed by space charge-limited conduction mechanism, which indicates the presence of an interfacial layer with imperfect microstructure near the top Al/CFO interface. The device shows nonvolatile behavior with good endurance properties, acceptable resistance ratio, uniform resistive switching due to stable, less random filament formation/rupture and a control over the resistive switching properties by choosing different stop voltages, which makes the device suitable for its application in future nonvolatile resistive random access memory (ReRAM).

The Resistive Random Access Memory (ReRAM) devices are potential candidate for the next generation non-volatile memory devices and has attracted considerable attention of scientific community due to low power consumption, high operation speed, excellent miniaturization potential, non-destructive readout, favourable scalability and good compatibility with complementary metal–oxide–semiconductor (CMOS) technology.[1–8] The ReRAM devices are based on resistive switching (RS) phenomena, in which the resistance of an insulating/dielectric material, sandwiched between two metallic electrodes, is changed by the application of externally applied electric field.[9] For many ReRAM devices initially an ''Electroforming'' process is usually required before the memory cell can start working. In this process the MIM structure is forced by an externally applied higher bias voltage to develop a conducting filament between top and bottom electrodes. But, a higher value of avalanche current can damage the switching device permanently. To prohibit this irreversible destruction of the device during the set process, it is necessary to apply a compliance current (CC), which makes an external current limiter essential in the ReRAM application that leads to a complicated circuit design of ReRAM based devices. Thus, it would be desirable to fabricate electroforming free ReRAM devices with self-compliance feature. Besides this, the commercialization of ReRAM devices is also hindered by several key issues of resistive switching, such as uniformity and reliability, and lack of control over the switching process, etc. It is necessary to pay much attention to get an electroforming free device with uniform resistive switching and self-compliance property.

$CoFe_2O_4$ is a stable oxide, which has been used for RS applications earlier also. Wang et al.[10] have used pulsed laser deposited CFO thin films in Pt/CFO/NSTO structure to study the RS properties of $CoFe_2O_4$. Munjal et al.[11] and Hu et al.[12] have used spin coated thin film to study the RS properties of $CoFe_2O_4$. RS in all the above mentioned studies suffer with the electroforming process and application of compliance current was found necessary. Using nanoparticles for RS applications may provide interesting results due to more oxygen vacancies or ion related defects at grain boundaries.

Herein, fabrication and study of an electroforming free, self-compliance Al/CoFe$_2$O$_4$/FTO ReRAM device using CoFe$_2$O$_4$ nanoparticles is reported with good and uniform resistive switching and controllability to obtain diverse stable and reliable resistance states by applying different stop voltages in reset process. CoFe$_2$O$_4$ nanoparticles' film (thickness ~400nm) was deposited on FTO substrate by spray coating the CoFe$_2$O$_4$ nanoparticles (size ~22nm), which were synthesized by hydrothermal method[13] using cobalt nitrate hexahydrate (0.05M) and ferric nitrate nonahydrate (0.1M) as starting precursors. The precursors were mixed thoroughly in water-ethanol solution under a vigorous stirring and 0.5M NaOH solution was added drop wise to the precursor solution to maintain the pH ~13. The final solution was transferred to a Teflon lined stainless steel autoclave and kept in a hot air oven preheated at 140 °C for hydrothermal treatment for 20 h. The structural characteristics of the synthesized CoFe$_2$O$_4$ nanoparticles were measured using a Philips X'Pert X-ray diffraction (XRD) system and surface morphology of CoFe$_2$O$_4$ nanoparticles thin films was observed by scanning electron microscopy (supporting information). The test device Al/CoFe$_2$O$_4$/FTO was obtained by depositing Al top electrodes (thickness ~100 nm, area 1 mm$^2$) by electron-beam evaporator using a shadow mask. Schematic representation of the fabricated resistive switching device is shown in the inset of Figure 1(a). The chemical state of top Al/CoFe$_2$O$_4$ interface was analysed by X-ray photoelectron spectroscopy. The I-V characteristic and resistive switching characteristic curves were obtained by Keithley-4200 SCS source-meter (sweep rate 1V/sec) using a LabVIEW program. In the test of Resistive Switching characteristics the voltage sweep was considered positive when the current flows from top electrode to bottom electrode.

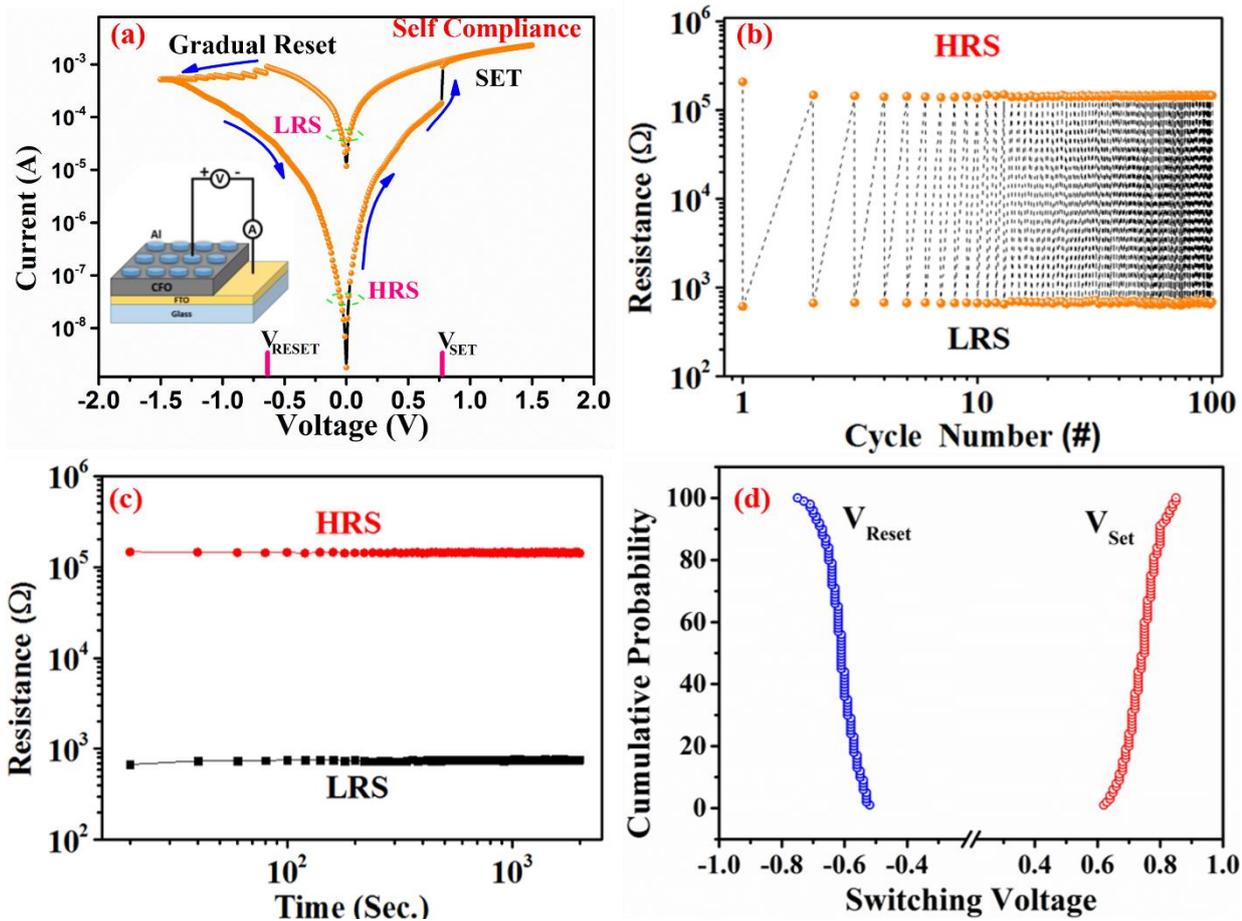

**Figure 1(a)** shows the resistive switching characteristic curve of the fabricated device. The arrows show the direction of voltage sweep. The inset in the figure shows the schematic representation of the fabricated device. Figure 1**(b)** depicts the program/erase endurance performance of the device, Figure 1**(c)** shows the retention properties of the fabricated device and 1**(d)** show the cumulative probability of the set and reset voltages.

Initially the device was in HRS and with a steady increase in the positive potential applied to the top electrode, a clear transition from the HRS to the LRS was observed at ~"0.78 V"; this transition is referred as the "SET" process. Subsequently, the opposite "RESET" process (at ~ "-0.6 V") was also observed when the device was subjected to a negative voltage sweep. We observed that the reset process in our device is a gradual process and device shows self-compliance feature during the set process, which indicates that the device does not need any compliance current (CC). In order to confirm the low power consumption capability of our RS device, we calculated the writing and erasing powers for set and reset processes in our device. For our RS device, the set and reset powers are 0.14 mW ($I_{set}$ = ~ 1.83×10$^{-4}$ A; $V_{set}$ = ~ 0.78 V) and 0.54 mW ($I_{reset}$ = ~ 9×10$^{-4}$ A; $V_{reset}$ = - 0.6 V), which are quite comparable to the reported powers of low power RS devices of other published works. For example, a low power RS device was fabricated by using ZnO nanorods, which uses ~ 9 mW power for reset and set process[14]. Lee et al.[15] has fabricated Pt/HfO$_x$/TiN/Si low power RS device with 0.25 and 0.15mW power for reset and set process.

Program/erase endurance test and data retention test were performed at room temperature for examining the switching reliability of Al/CoFe$_2$O$_4$/FTO ReRAM device. In order to perform the program/erase endurance tests, the I-V measurements of the device in the cyclic voltage sweep are repeated 100 times and the resistance in HRS and LRS at read voltage of 0.1 volt with cycle number is shown in Figure 1(b). The endurance performance of the fabricated device demonstrates a very good bipolar resistive switching over 100 cycles and we did not observe any noticeable degradation in the performance. For data retention test in LRS a continuous read voltage of 0.1 V was applied and stability of the resistance state with time is observed. In addition, for retention test in HRS the device is switched OFF and test was performed. During the data retention tests, the device maintained high resistance state (HRS) or low resistance state (LRS), for >10$^3$ sec without any noticeable fluctuation and the resistance ratio is always >10$^2$ (Figure 1(c)), which confirms the nonvolatile behavior of the fabricated device. We plotted the cumulative probability of the set and reset voltage distribution of the device for studying the device resistive switching stability (Figure 1(d)). The fabricated device shows very stable switching behavior for different RS memory cycles and cells (supporting information).

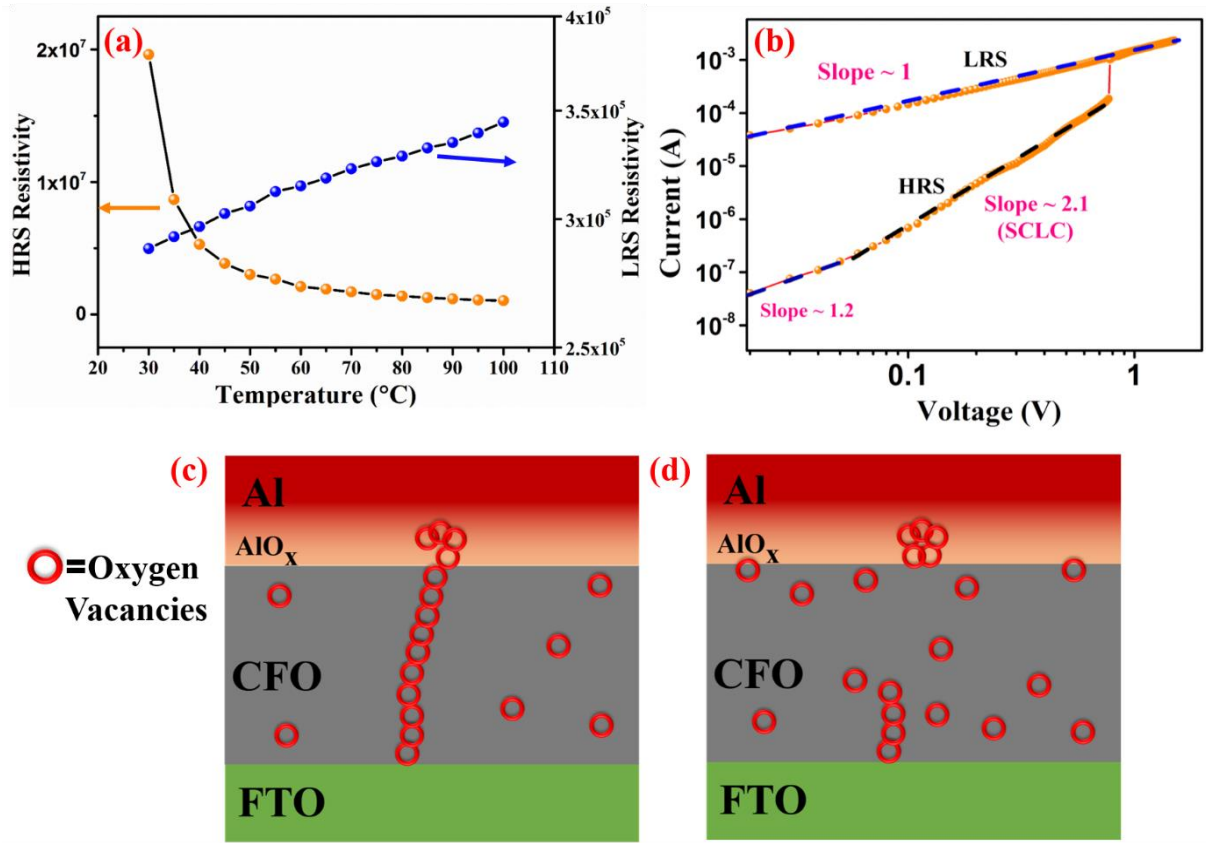

**Figure 2(a)** Resistance vs. Temperature curves of the device in HRS and LRS. 2**(b)** Linear fitting of I–V characteristic curve of Al/CoFe$_2$O$_4$/FTO device in positive bias for HRS and LRS. 2**(c)** and 2**(d)** the low resistance state and high resistance state of the device with a stable conducting path in interfacial AlO$_x$ layer.

In order to understand the conduction mechanisms and resistive switching of our fabricated Al/CoFe$_2$O$_4$/FTO device, temperature dependence of the resistance (R-T) measurements of the device and I-V characteristics were studied when the device was in low resistance state or in high resistance state. Figure 2 (a) shows the results of the R-T studies (applied bias = 0.1 V) in the temperatures range from room temperature to 100°C. In the low resistance state the device exhibit metal like behavior whereas in the high resistance state a semiconducting behavior was observed. To obtain the value of resistance temperature coefficient of (α), we used R(T)= R$_0$[1+α(T–T$_0$)] relation to linearly fit the temperature dependent resistance data. Linear fitting gives α = 0.0021 K$^{-1}$. This value of α is in good agreement with previously reported values of resistance temperature coefficient for the oxygen vacancy filaments in resistive switching devices[16,17]. By fitting the temperature dependence of resistance (R) in HRS to the Arrhenius equation $R=R_0 \cdot e^{-\Delta E/kT}$, where $R_0$ is the resistance at room temperature, $\Delta E$ is the thermal activation energy of charge carriers, k is the Boltzmann constant, and T is the temperature in Kelvin. Here, $\Delta E$ was estimated as ~62 meV. The electrons interacting with the shallow oxygen defects are generally thermally activated so that band conduction is dominant.

We fitted the I-V curves of several RS cycles as well as different RS memory cells and in Figure 2(b) the fitting of a typical RS cycle is shown. The lnI–lnV curve in the LRS state shows a slope ~1, which indicates an Ohmic behavior and can be attributed to the formation of a metallic type conducting path in the switching device during the SET process, which is in agreement with the R-T studies. For HRS, in small voltage regime (<0.07 V ($V_1$)), the slope of the fitted line in the ln(V) vs ln(I) curve was ~1.2; which indicates the presence of Ohmic type conduction.[18] In this regime the conduction is mainly dominated by thermally excited electrons and the injected electrons are not comparable with intrinsic thermally excited electrons. However, in HRS under comparatively higher bias (>0.07 V ($V_1$)), the slope of curve increases to ~2.1 and in this regime is space charge limited current (SCLC) type conduction dominates.[19] The nature of the I-V characteristics remains similar for different RS cycles except a small variation in $V_1$. The SCLC conduction is obtained if the electrode's contact is highly carrier injective, which can occur if an interfacial layer of metal oxide gets formed.[20] In our device, we have Al as top electrode, which is a readily oxidizable material and can provide an oxide interfacial layer ($AlO_x$) at the $Al/CoFe_2O_4$ interface.

We fabricated a new $Al/CoFe_2O_4/FTO$ device with top contact area ~$10^{-2}$ $mm^2$. The contact pad area of this device was much smaller as compared to the earlier studied device (contact pad area ~ 1$mm^2$). The new device also showed electroforming free resistive switching and the set and reset voltage were ~ 0.76 V ($V_{set}$) and ~ -0.57 ($V_{reset}$) respectively (supplementary material). The resistance of LRS and HRS were ~ 0.7 k$\Omega$ and 142 k$\Omega$ respectively. The values of $V_{reset}$ and $V_{set}$ voltages and resistance ratio for the Al/CoFe$_2$O$_4$/FTO device with contact pad area ~10$^{-2}$ mm$^2$ are almost similar to the device of contact pad area 1 mm$^2$. Switching voltage and resistance of LRS as well as HRS remains almost similar even on decreasing the contact area significantly, indicates that the switching occurs locally in the CoFe$_2$O$_4$ film through a filamentary conducting path.[21] Similar results have been observed by Chen et al.[22], where the authors have found no dependence of the top contact area on the switching voltage in the filamentary type resistive switching. The observed RS in Al/CoFe$_2$O$_4$/FTO device can be explained by considering the formation/rupture of conducting filaments. Figure 2(c) and 2(d) shows the model for "ON" and "OFF" state. The stable RS behaviour seems to be due to the formation of stable conducting filament in AlO$_X$ layer. When positive voltage is applied on the top Al electrode, the oxygen vacancies from the Al/CoFe$_2$O$_4$ interface moves toward the bottom electrode, which is equivalent to the migration of oxygen ions (O$^{2-}$) to the Al/CoFe$_2$O$_4$ interface and accumulate near the top electrode, leaving behind a conducting path of oxygen vacancies that directly connects the bottom FTO electrode to the top Al electrode and the device switches from HRS to LRS. On applying the negative bias to the top electrode the reverse action may occur, leading to a dissolution of the conducting filament, which takes the device back to HRS.

In order to confirm the presence of oxygen deficient interfacial layer, the chemical state of Al near the Al/CoFe$_2$O$_4$ interface was also analyzed by X-ray photoelectron spectroscopy (XPS) after sputter-etching the top electrode (etching rate ~4 nm/min) with Ar$^+$ ions (2kV, emission current 2 µA). The aluminum near the interface exhibits mixed oxidation states of Al$^0$, Al$^{1+}$, Al$^{2+}$ and Al$^{3+}$ (Figure 3(a)) and confirms the presence of a mixed oxide AlO$_x$ interfacial layer. Such mixed oxides naturally have an imperfect microstructure, and metallic ions or oxygen vacancies are likely to exist in the matrix.[23] The oxygen-deficient AlO$_x$ phase has been reported to provide the electron trapping sites[24] and can be a reason behind the SCLC conduction observed in our device as observed during I-V measurements in HRS.

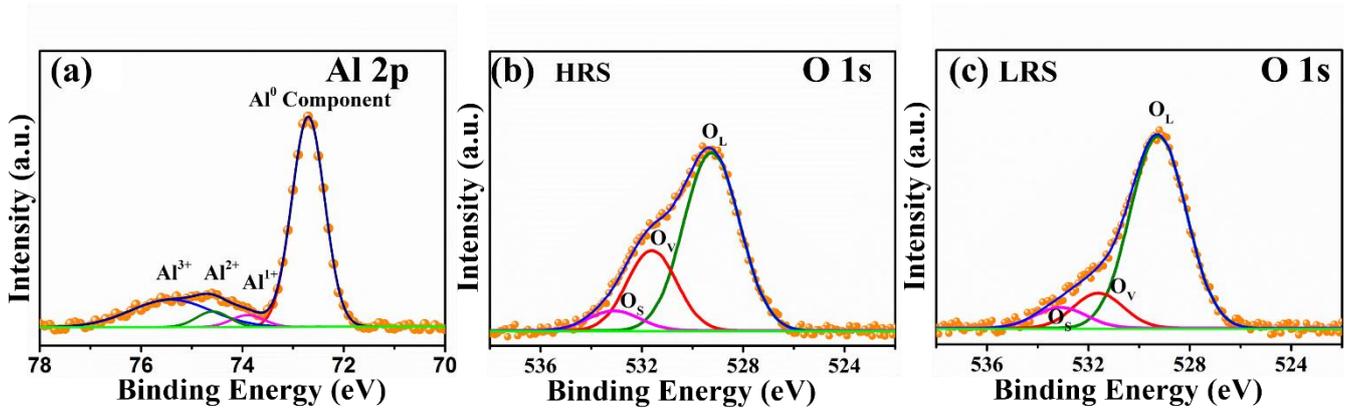

**Figure 3** XPS spectra of **(a)** Al 2p core level, **(b)** O1s core level (HRS) and **(c)** O1s core level (LRS) at the Al/CoFe$_2$O$_4$ interface of the device.

In our fabricated RS device, the presence of self-compliance feature may also be attributed to the formation of AlO$_x$ layer near the top electrode between Al and CFO. The presence of a thin AlO$_x$ layer near the top electrode of the resistive switching device behaves like a built-in series resistance, which reduces the operational current and gives a stable self-compliance property[25].

The O1s XPS spectra near the Al/CoFe$_2$O$_4$ interface indicates the presence of lattice oxygen, oxygen vacancies and surface oxygen. Electroforming free RS behaviour observed in our device can be attributed to these oxygen vacancies as electroforming free behaviour has been observed earlier also in RS devices containing a high defect concentration[18,19,26–28]. Similar behaviour has been observed in the case of resistive switching device fabricated using titanium oxide nanoparticles, where the authors have suggested that the fast diffusion of oxygen vacancies at the boundaries between the nanoparticles can be the reason of electroforming free resistive switching.[29] Lanza et al.[30] have also shown that the grain boundaries shows higher conductivity comparing to that of the grains is related to an excess of the oxygen vacancies at the grain boundaries in oxide based resistive switching device. Figure 3(b)-(c) shows the XPS spectra of O1s core level at the Al/CFO interface for the HRS and LRS. It is clear from the O 1s XPS measurements that the oxygen vacancies' concentration is higher when the device is in HRS and decreases significantly when the device switches to LRS. The Oxygen vacancies are positively charged sites and are formed near the top Al electrode in our device. So, when the device is in HRS more oxygen vacancies are present near the Al/CFO interface. On applying the positive bias on the top electrode these positively charged oxygen vacancies are pushed away from the interface to form a conducting filament connecting the top and bottom electrodes and the device switches from HRS to LRS. This process is expected to decrease the concentration of oxygen vacancies near the top electrode, as observed in the XPS measurements (Figure 3 (c)). On changing the polarity of bias voltage applied to the top electrode the positively charged oxygen vacancies are pulled back to interface which breaks the conducting filament and disconnects the bottom electrode from top electrode. This process switches the device again to HRS from LRS, and the concentration of oxygen vacancies near the top electrode is expected to increase, which agrees with the observed XPS results for the device in HRS (Figure 3 (b)).

In the present Al/CoFe$_2$O$_4$/FTO device, we have also observed that different stop points in a gradual reset process of the device leads to achieve diverse HRS levels as shown in Figure 4(a). The observed different HRS may be attributed to different proportion of filament dissociated under different negative electric field (-0.9V, -1.2V and -1.5V), which results in completely distinguished high resistance states. When reset stop point is smaller (-0.9V), it dissolves only a small amount of filament, which results in a small resistance change. On applying a comparatively higher negative stop voltage (-1.2V or -1.5V), relatively more fraction of the filament is dissolved and consequently more change in resistance is achieved while switching from LRS to HRS (Figure 4 b-d).

The controllability over the RS behavior confirms the presence of a stable conducting channel in the AlO$_X$, which improves the uniformity of resistive switching by nucleating the regrowth of conducting filaments in CoFe$_2$O$_4$ layer during the next set process and bring the controllability over filament, as well as it gives the ability to modify resistive switching behavior.

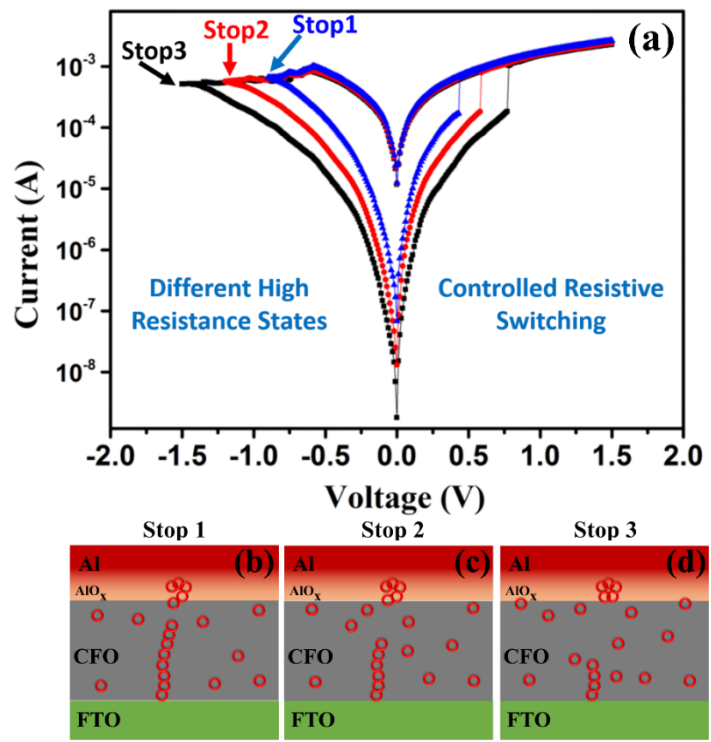

**Figure 4(a)** I–V curves of the Al/CoFe$_2$O$_4$/FTO RS device measured at different reset stop voltages to confirm the controllability of the filament. **(b–d)** Schematic representation of resistive switching mechanism for 3 different HRS.

In summary, we fabricated a reliable, electroforming free, self-compliance resistive switching device with Al/CoFe$_2$O$_4$/FTO structure. The fabricated device shows good data retention properties and program/erase endurance properties. A model based on the formation of stable conducting channel in AlO$_X$ serving as a nucleation point, from which the conducting filament formed in CoFe$_2$O$_4$ layer tend to grow in the subsequent switching cycles is proposed to explain the RS behavior. The presence of AlO$_X$ layer was confirmed by the XPS studies. The improved uniformity or reduced randomness in our device may be attributed to low voltage and low current operation as well as less random filament formation and rupture.

See supplementary material for more details about the x-ray diffraction, the scanning electron microscopy and RS operational uniformity.

The financial support from MeitY (Government of India) is gratefully acknowledged. One of us (SM) is also thankful to Council of Scientific and Industrial Research (CSIR), New Delhi for senior research fellowship (SRF) Grant.us (SM) is also thankful to Council of Scientific and Industrial Research (CSIR), New Delhi for senior research fellowship (SRF) Grant.